# Deterministic control of photonic de Broglie waves using coherence optics: Coherence de Broglie waves

Byoung S. Ham

Center for Photon Information Processing, School of Electrical Engineering and Computer Science, Gwangju Institute of Science and Technology

123 Chumdangwagi-ro, Buk-gu, Gwangju 61005, S. Korea

(Submitted on February 1, 2020)

bham@gist.ac.kr

In quantum mechanics, photonic de Broglie waves have been understood as a unique property of quantum mechanics satisfying the complementarity between particle and wave natures of light, where the photonic de Broglie wavelength is inversely proportional to the number of entangled photons acting on a beam splitter. Very recently, the heart of nonclassical feature of photon bunching on a beam splitter was newly interpreted using pure wave nature of coherence optics [arXiv:1911.07174v2], paving a road to unconditionally secured classical key distribution [arXiv:1807.04233v3]. Here, Mach-Zehnder interferometer-based deterministic photonic de Broglie waves are studied in a coherence regime for both fundamental physics and potential applications of coherence-quantum metrology.

The nonclassical feature of anticorrelation on a beam splitter (BS), the so-called Hong-Oh-Mandel dip or photon bunching, has been the heart of quantum mechanics in terms of superposition and entanglement, where it cannot be achieved by classical means[1-5]. Unlike most anticorrelation studies based on statistical nature of light, a deterministic solution has been recently found in a coherence manner for a particular phase relation between two input fields impinging on a BS[6]. Owing to coherence optics with a phase control, the BS-based anticorrelation can be achieved in a simple Mach-Zehnder interferometer (MZI)[6]. One of the first proofs of the MZI physics for quantum mechanics was for anticorrelation using single photons[1]. In the coherence version, unconditionally secured classical key distribution has been proposed[7]. Although the physics of the unconditionally secured classical key distribution is based on quantum superposition, i.e., indistinguishability in the MZI paths[7], the key carrier is not a quantum but a coherent light compatible with current fiber-optic communications networks. As debated for several decades, a fundamental question about quantum nature of light is still an on-going important subject in quantum optics community[8-10].

Here in this paper, a fundamental question of "what is the quantum nature of light? or "what is the origin of nonclassicality?" is asked and answered in terms of photonic de Broglie waves (PBW) in a pure coherence regime based on the wave nature of light. Due to the quantum property of linear optics such as a BS or MZI, however, the nonclassical light is also included to the present scope. Thus, the present paper is for general conceptual understanding in fundamental quantum physics as well as for potential applications of coherence-quantum metrology to overcome the statistical quantum limitations such as an extremely low rate at the higher-order entangled photon-pair generation.

The photonic de Broglie wavelength $\lambda_B$ has been a key feature in quantum mechanics of wave-particle duality or complementarity for quantum nature of light, where classical physics has been completely blocked off[11-14]. The PBW has been demonstrated using entangled photon pairs generated from spontaneous parametric down conversion (SPDC) process, where $\lambda_B = \lambda_0/N$, and $\lambda_0$ (N) is the initial wavelength (number of entangled photons in such as a NOON state) of light[11-14]. For example, a single-photon entangled pair on a beam splitter results in PBW at $\lambda_B = \lambda_0/2$. So does $\lambda_B = \lambda_0/4$ for a two-photon entangled pair. Due to experimental difficulties of obtaining a higher-order entangled photon pair, however, the application of quantum PBW has been severely limited so far, whose latest record is $\lambda_B = \lambda_0/18$ with N=18[14]. By the same reason, quantum metrology such as quantum lithography and quantum sensing has also been limited to practical applications[15-17]. Most of all, there is no deterministic entangled photon pair generator, yet.

In the present paper, a deterministic control of PBW using the coherence-optics-based anticorrelation[6] is presented for both fundamental physics and its potential applications of coherence-quantum metrology, where the order N in $\lambda_B$ is potentially unlimited and on demand. The deterministic control of PBW should give a great benefit to quantum metrology beyond the standard quantum limit. The deterministic controllability of the higher-order PBW brings a breakthrough in practical limitations of the entangled photon-based conventional quantum metrology[15-17]. Most of all, understanding of the quantum nature of light in PBW is the most important result. For this a cross coupled double (CCD) MZI scheme is used for a typical laser.

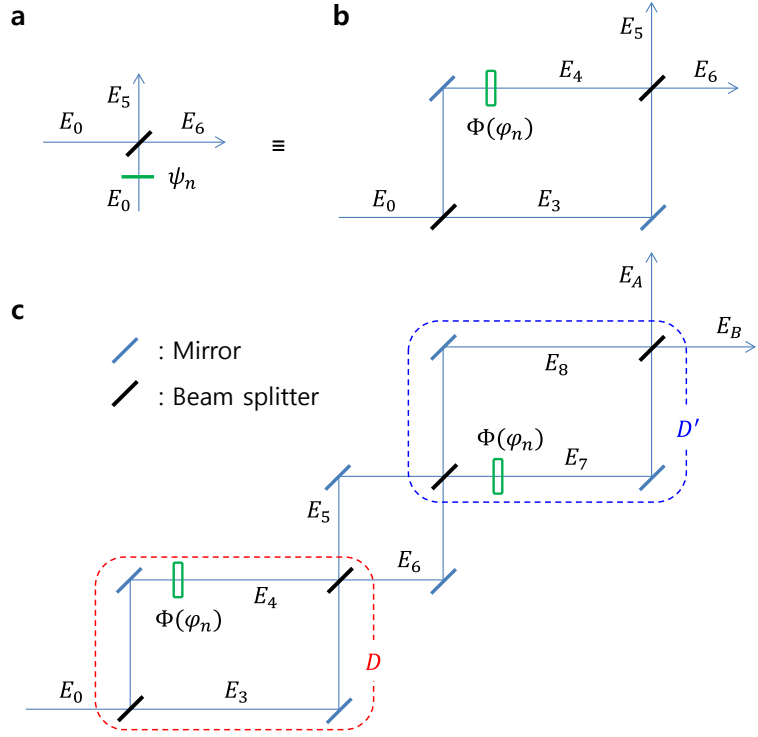

**Figure 1.** A basic unit of coherence PBW. **(a)** A BS-based anticorrelation scheme for photon bunching. **(b)** An equivalent scheme of **(a)**. **(c)** A basic unit of coherence PBW. The input field $E_0$ is coherent light. D or D' indicates a MZI building block composed of beam splitters and a phase shifter. The coupled matrix of $[D'][D]$ represents a coherence PBW scheme equivalent to quantum PBW with N=4.

Figure 1 shows a basic building block of the present coherence PBW using coherence optics-based anticorrelation. Figure 1(a) shows a deterministic scheme of anticorrelation with a phase shifter $\psi_n$ for photon bunching or a HOM dip on a BS[6]. The controlled phase $\psi_n$ is to clarify the statistical single photon-based anticorrelation[1-5], where the anticorrelation on a BS must suffice the magic phase between two input photons: $\psi_n = \pm(n-1/2)\pi$ and n=1,2,3…[6]. Thus, the vagueness in conventional anticorrelation on a BS has been clearly understood and extended into a deterministic feature of nonclassical light generation. Because BS matrix satisfies a π/2 phase shift between two split outputs, i.e., reflected and transmitted lights[18], Fig. 1(a) can be simply represented by a typical MZI as shown in Fig. 1(b)[1]. Due to the preset π/2 phase shift on the first BS for $E_3$ and $E_4$ in Fig. 1(b), the inserted phase shifter of $\varphi_n$ must be $\varphi_n = \pm n\pi$ for the same outputs as in Fig. 1(a)[6]. The intensity correlation g[(2)] between two outputs $I_5$ and $I_6$ is described by $g^{(2)} = \frac{\langle I_5 I_6 \rangle}{\langle I_5 \rangle \langle I_6 \rangle}$, where $I_j$ is the intensity of $E_j$. Thus, conventional MZI becomes a quantum device for nonclassical photon generation with determinacy for a Schrodinger's cat or a NOON state[1,6].

In the conventional photon bunching phenomenon as shown in a HOM dip or a Bell state using SPDC-based entangled photon pairs, the requirement of $\psi_n$ in Fig. 1(a) is automatically satisfied by a closed-type $\chi^{(2)}$−based three-wave mixing process in a nonlinear medium. In the SPDC nonlinear optical process, however, the choice of the sign of $\psi_n$ cannot be deterministic due to the bandwidth-wide, probabilistically distributed space-superposed entangled photons as described by, e.g., a polarization entanglement superposition state[2]: $|\psi\rangle = \left(|H\rangle_1|V\rangle_2 + e^{i\psi}|V\rangle_1|H\rangle_2\right)/\sqrt{2}$. In the case of two independent solid-state emitters, the generated single photon pair must be phase-locked if excited by the same pump pulse. Thus, the condition of $\psi_n$ in Fig. 1(a) must be postadjusted to be $\pm\frac{\pi}{2}$ in the relative phase difference for the anticorrelation or an entangled state generation[3]. The proof of the magic phase of $\pm\frac{\pi}{2}$ in Fig. 1(a) for nonclassical light generation has already been demonstrated in two independent trapped ions[19]. In Fig. 1(b), the spectral bandwidth ($\delta\omega$) of the input light $E_0$ should limit the interaction time ($\tau$) or coherence length ($l_C$) in $g^{(2)}$ anticorrelation. In the application of secured communications[7], the transmission distance is potentially unlimited, where $l_C = \frac{c}{mHz} \sim 10^8 \ (km)$ if sub-mHz linewidth laser is used[20]. In this case, a common phase encoding technique may be advantageous compared to the amplitude modulation technique. According to ref. 21, the maximal indistinguishability induced by perfect quantum superposition represents for maximal coherence, where maximal coherence is a prerequisite for an entangled state or nonclassical light generation.

Figure 1(c) represents a basic building block of the present CCD-MZI for a deterministic control of PBW via coherence optics-based anticorrelation. The output fields in the first building block D of Fig. 1(c), whether it is for $E_5$ or $E_6$, are fed into the block D' by splitting into $E_7$ and $E_8$, resulting in the second-order superposition state. Here, the condition of anticorrelation in MZI is $\varphi_n = \pm n\pi$, resulting in a distinctive output either $E_5$ or $E_6$. The same phase shifter is used and simultaneous controlled in both D and D' in an asymmetric configuration: see the phase shifter $\Phi(\varphi)$ locates oppositely in each block. If the phase shifter distribution is symmetric, then a unitary transformation is applied for the unconditionally secured classical cryptography[7]. The second-order superposition in Fig. 1(c) offers the fundamental physics of the present coherence PBW. The output of the first block D in Fig. 1(c) is described as follows:

$$\begin{bmatrix} E_5 \\ E_6 \end{bmatrix} = [D]\begin{bmatrix} E_0 \\ 0 \end{bmatrix} = \frac{1}{2}\begin{bmatrix} 1 - e^{i\varphi} & i(1 + e^{i\varphi}) \\ i(1 + e^{i\varphi}) & e^{i\varphi} - 1 \end{bmatrix}\begin{bmatrix} E_0 \\ 0 \end{bmatrix}, \tag{1}$$

where $[D] = [BS][\Phi][BS]$, $[BS] = \frac{1}{\sqrt{2}}\begin{bmatrix} 1 & i \\ i & 1 \end{bmatrix}$, and $[\Phi] = \begin{bmatrix} 1 & 0 \\ 0 & e^{i\varphi} \end{bmatrix}$. As already known in the MZI interferometry, equation (1) shows a $2\pi$ modulation period in each output intensity: $I_5 = I_0(1 - \cos(\varphi))$; $I_6 = I_0(1 + \cos(\varphi))$ as shown in Fig. 2(a). Thus, the intensity correlation $g^{(2)}$ has a $\pi$ modulation as expected (see the red curve in Fig. 2(a)):

$$g_{56}^{(2)} = [1 - \cos(2\varphi)]/2, \tag{2}$$

where the phase basis for $g_{56}^{(2)} = 0$ is $\varphi_n = \pm n\pi$. This is a new understanding of the nonclassical feature, where a perfect coherence plays a major role. Equation (2) is known as the classical resolution limit or Rayleigh criterion[22].

The output lights, $E_A$ and $E_B$, in the second block D' of Fig. 1(c) are then described by the following relation:

$$\begin{bmatrix} E_A \\ E_B \end{bmatrix} = [D'][D]\begin{bmatrix} E_0 \\ 0 \end{bmatrix} = -\frac{1}{2}\begin{bmatrix} 1 + e^{i2\varphi} & i(1 - e^{i2\varphi}) \\ -i(1 - e^{i2\varphi}) & 1 + e^{i2\varphi} \end{bmatrix}\begin{bmatrix} E_0 \\ 0 \end{bmatrix}, \tag{3}$$

where $[D'] = [BS][\Phi'][BS]$ and $[\Phi'] = \begin{bmatrix} e^{i\varphi} & 0 \\ 0 & 1 \end{bmatrix}$. Unlike equation (1), equation (2) results in a twice shorter (faster) modulation period (frequency), i.e., $\pi/2$ modulation in each intensity of $I_A$ and $I_B$: $I_A = 2(1 + \cos(2\varphi))$; $I_B = 2(1 - \cos(2\varphi))$ (see Fig. 2(b)). As a result, the intensity correlation $g^{(2)}$ of $I_A$ and $I_B$ in Fig. 1(c) becomes:

$$g_{AB}^{(2)} = [1 - \cos(4\varphi)]/2. \tag{4}$$

Thus, the classical resolution limit of $\lambda_0/2$ governed by the Rayleigh criterion in Fig. 1(b) is overcome now using coherence optics in Fig. 2(b). This is the first proof in history for the nonclassical feature obtained by pure coherence optics. In equation (4), the phase basis for $g_{AB}^{(2)} = 0$ is accordingly changed from $\varphi_n = \pm n\pi$ in Fig. 1(b) to $\varphi_n = \pm n\pi/2$. This doubly enhanced resolution of the output intensity in Figs. 1(c) and 2(b) should contradict to the general understating of quantum mechanics because the method of Fig. 1(c) and its result in Fig. 2(b) are perfectly coherent and classical.

Here, it should be noted that the perfect coherence between two lights ($E_3/E_4$ or $E_7/E_8$) is achieved from path indistinguishability in MZI, satisfying the anticorrelation condition in the outputs ($E_5/E_6$ or $E_A/E_B$)[6]. Thus, the specific phase relation with $\varphi_n$ between two superposed coherent lights in MZI of Fig. 1(b) becomes the definite source of nonclassical feature such as anticorrelation and entanglement[6]. In that sense, the number of superposition state in Fig. 1(c) should be equivalent to the number of entangled photons in conventional quantum PBW (see equations (2) and (4)). Therefore, the CCD-MZI scheme composed of D and D' in Fig. 1(c) represents the basic unit of the present coherence version of PBW. Then, the higher-order coherence PBW can also be achieved by simply connecting the asymmetrical unit of Fig. 1(c) in a series (discussed in Fig. 3).

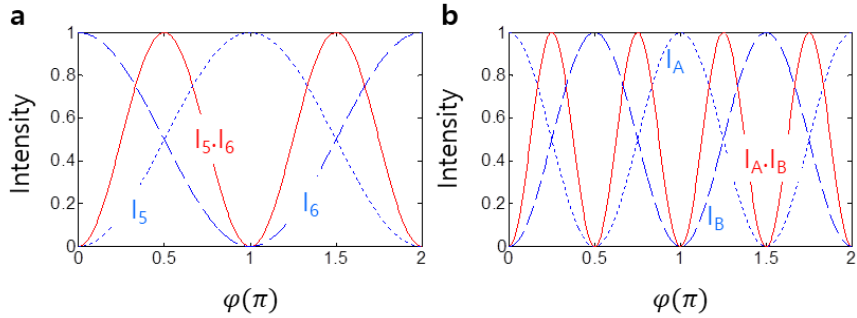

**Figure 2.** Numerical calculations for $g^{(2)}$ intensity correlation of Fig. 1(c). **(a)** Red: $I_5 I_6$ (normalized), Dotted: $I_5$, Dashed: $I_6$. **(b)** Red: $I_A I_B$ (normalized), Dotted: $I_A$, Dashed: $I_B$. The input field intensity of $E_0 = 1$ is assumed.

Figure 2 shows numerical calculations for Fig. 1(c) to support the present theory of the deterministic control of PBW in a coherence regime. Figure 2(a) shows a typical MZI result of Fig. 1(b) by solving equation (1), where each output intensity represents the classical limit. As expected, the conventional MZI scheme gives a spectroscopic resolution of $\lambda_0/2$, which is the Rayleigh limit in classical physics. This classical resolution limit is now understood as the first order of the present deterministic control of PBW: $\lambda_{CB} = \lambda_0/2\zeta$, where $\zeta$ is the number of MZI block (or superposition state in the form of Fig. 1(b)), and $\lambda_{CB}$ indicates the present coherence PBW. Here, it should be noted that each MZI block in Fig. 1(b) is equivalent to N=2 in quantum PBW for an entanglement superposition description at $|\psi\rangle = (|N\rangle_A|0\rangle_B + |0\rangle_A|N\rangle_B)/\sqrt{2}$: $2\zeta = N$. In other words, a typical MZI is a quantum device for anticorrelation or nonclassical light generation if $\varphi_n = \pm n\pi$ is satisfied. The intensity correlation of $g_{AB}^{(2)}$ in equation (4) is numerically calculated in Fig. 2(b): see red curve. The demonstration of $\lambda_{CB} = \lambda_0/4$ in Fig. 2(b) proves the present theory of coherence PBW based on Fig. 1(c). Thus, it is concluded that the present coherence PBW in Fig. 1(c) is equivalent to the quantum PBW based on entangled photons with an additional benefit of deterministic controllability.

For the higher order $\lambda_{CB}$, the basic scheme of Fig. 1(c) needs to be repeated in a series or circulated in a feedback form as shown in Fig. 3(a). In the serial connection, the output from each block becomes an input to the next block without loss. Defining $[CM] = [D'][D]$, the n$^{th}$ order output fields in Fig. 3(a) can be obtained from equation (3):

$$\begin{bmatrix} E_A \\ E_B \end{bmatrix}^n = [CM]^n \eta^{2(n-1)} \begin{bmatrix} E_0 \\ 0 \end{bmatrix}, \quad (5\text{-}1)$$

$$= \frac{1}{2}(-1)^n \eta^{2(n-1)} \begin{bmatrix} (1 + e^{i2n\varphi}) & i(1 - e^{i2n\varphi}) \\ -i(1 - e^{i2n\varphi}) & (1 + e^{i2n\varphi}) \end{bmatrix} \begin{bmatrix} E_0 \\ 0 \end{bmatrix}, \quad (5\text{-}2)$$

$$(E_A)^n = \frac{E_0}{2}(-1)^n \eta^{2(n-1)}(1 + e^{i2n\varphi}), \qquad (5\text{-}3)$$

$$(E_B)^n = i\frac{E_0}{2}(-1)^{n+1}\eta^{2(n-1)}(1 - e^{i2n\varphi}). \qquad (5\text{-}4)$$

From equations (5-3) and (5-4) the related n$^{th}$ order intensities are obtained:

$$(I_A)^n = \frac{1}{2}I_0[1 + \cos(2n\varphi)], \qquad (6\text{-}1)$$

$$(I_B)^n = \frac{1}{2}I_0[1 - \cos(2n\varphi)]. \qquad (6\text{-}2)$$

where $I_0 = E_0 E_0^*$. Regardless of the chain length the final output intensity is always the same as the input. As a result, the n$^{th}$ order intensity correlation $g_n^{(2)}$ becomes:

$$g_n^{(2)} = \frac{\langle (I_A)^n (I_B)^n \rangle}{\langle (I_A)^n \rangle \langle (I_B)^n \rangle} = \frac{1}{2}[1 - \cos(4n\varphi)]. \qquad (7)$$

Thus, the general solution for the the n$^{th}$ order coherence PBW in Fig. 3 is:

$$\lambda_{CB}^{(n)} = \lambda_0/4n, \qquad (8)$$

where n is the number of blocks or CCD-MZI. For n=1, there are basic building blocks of D and D' equivalent to the four-photon (N=4) case in quantum PBW[11-14]. Because equation (8) is deterministic, the present coherence PBW is powerful compared with conventional quantum counterpart in terms of N number as well as determinacy and controllability. These facts may open a door to coherence-quantum metrology based on on-demand $\lambda_{CB}^{(n)}$. For the impracticality of quantum PBW, it takes ~2 hour acquisition time just for N=18[14].

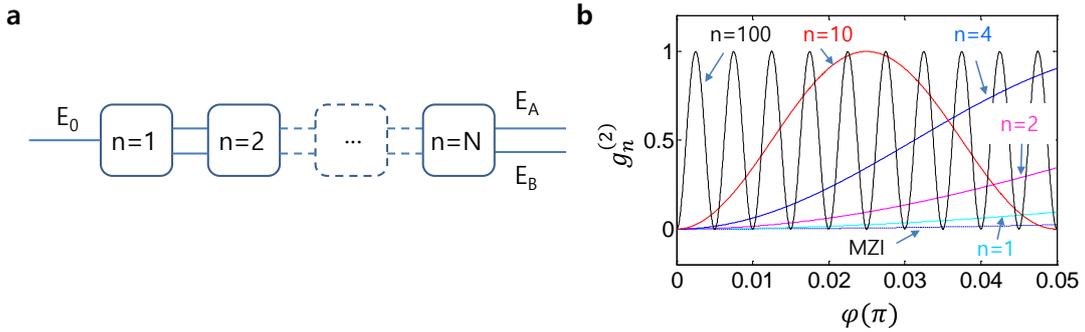

**Figure 3.** A photonic de Broglie wavelength generator. **(a)** A serially connected CCD-MZI. Each block represents CCD-MZI of Fig. 1(c). **(b)** Numerical calculations for **(a)**, where n indicates the number of blocks in **(a)**. MZI represents a reference of a classical limit whose period is $\pi$ as shown in Fig. 2(a).

Figure 3(a) represents a serially connected CCD-MZI scheme, where each block is equivalent to CCD-MZI of Fig. 1(c) at four-photon quantum PBW: $\lambda_B \ (= \frac{\lambda_0}{4})$. For the connection there are two lines, where only one line is active if the anticorrelation condition $\varphi_n = \pm n\pi$ is satisfied. An error is indicated in the ouput intensity loss via decoherence resulting in both lines active. The error sharpness is the spectroscopic resolution for coherence-quantum metrology. To realize Fig. 3(a), a waveguide[23] or a fiber-coupled[24] MZI scheme would be a good example.

Figure 3(b) shows numerical calculations using equation (7) for the intensity correlation $g_n^{(2)}$. As shown, the coherence $\lambda_{CB}^{(n)}$ is clearly equivalent to the quantum $\lambda_B$. Compared with quantum PBW[12-14], the present coherence PBW at $\lambda_{CB}^{(n)}$ is deterministic, real time, and virtually no limit in n. Each intensity modulation period for $(I_A)^n$ or $(I_B)^n$ is, of course, twice longer than $g_n^{(2)}$, as shown in equations (6-1) and (6-2).

In conclusion, the deterministic control of photonic de Broglie waves (PBW) was presented in a purely coherence manner for both fundamental physics and potential applications of coherence-quantum metrology using a chain of CCD-MZI scheme. For this, the output from each CCD-MZI was directly inserted to the next one until the given n number reaches, where n is the number of CCD-MZI. The analytical expression and its numerical calculations successfully demonstrated an equivalent feature to the quantum PBW, where number of MZIs in the present coherence PBW is equivalent to the entangled photon number N in quantum PBW. The random phase noise of the MZI system caused by mechanical vibrations, air turbulence, and temperature variations at ≤MHz rate may be eliminated by the use of either silicon photonics or fiber technology. The input light is the limit of coherence PBW in the MZI scheme, a fine laser such as sub-mHz laser gives a higher n number or shorter PBW[25]. As a result, present coherence PBW can be directly applied to high precision optical spectroscopy or quantum metrology such as optical clock[26], gravitational wave detection[27], quantum lithography[15,16], and quantum sensors[17]. The seemingly contradiction of coherence PBW to quantum physics is reconciled by quantum superposition of MZI paths, where MZI is treated as a quantum device like BS if the magic phase is involved[1,6]. Thus, the present scheme of Fig. 3(a) may open a door to coherence-quantum metrology for deterministic control of photonic de Broglie wavelength at higher orders in real time and for on-demand. Eventually, the present CCD-MZI-based photonic de Broglie wave generation scheme may apply for non-classical light generation such as deterministic entangled photons and photonic qubits, resulting in on-demand quantum information processing (discussed elsewhere).

Author's note: Arxiv:2001.06913v2/v3 contains some mistakes in analysis added to v1. This version is the updated v1.